\documentclass[]{aa}
\usepackage{graphics}
\usepackage{natbib}
\usepackage{longtable}
\bibpunct{(}{)}{;}{a}{}{,} 

\usepackage{txfonts}
\newcommand{\s}{\scriptsize}
\begin{document}
\title{7~mm continuum observations of ultra compact HII regions}
\author{
P. Leto\inst{1,}\inst{2}
G. Umana\inst{1},
C. Trigilio\inst{1},
C. S. Buemi\inst{1},
S. Dolei\inst{3},
P. Manzitto\inst{3},
L. Cerrigone\inst{4},
C. Siringo\inst{3}
}

\offprints{P. Leto \email{paolo.leto@oact.inaf.it}}

\institute{
INAF - Osservatorio Astrofisico di Catania, Via S. Sofia 78, 95123 Catania, Italy
\and
INAF - Istituto di Radioastronomia, Sezione di Noto, C.P. 161, Noto (SR), Italy
\and 
Universit\`a  di Catania, Dipartimento di Fisica e Astronomia, Via S. Sofia 78 , 95123 Catania, Italy
\and 
Max-Planck-Institut f{\"u}r Radioastronomie, Bonn, Germany
}

%\date{Received date, accepted date }

\titlerunning{7~mm continuum observations of ultra compact HII regions}
\authorrunning{P. Leto et al.}

\abstract  
% context heading (optional)
{} %leave it empty if necessary  
% aims heading (mandatory)
{
Ultra compact HII (UCHII) regions are indicators of high-mass star formation sites
and are distributed mainly in the Galactic plane. They exhibit a broad band spectrum with significant
emission between near-IR and radio wavelengths.
We intend to investigate the possible contribution of the forthcoming ESA Planck mission to the science
of UCHII regions by evaluating the possibility of detecting 
UCHIIs that are bright in the radio regime.
}
% methods heading (mandatory)
{
We performed new 7~mm observations of a sample of UCHII regions.
The observations were designed 
to acquire high-frequency radio spectra. 
For each source in our sample, the free-free radio spectrum has been  modeled.
Along with far-IR measurements, 
our spectra allow us to estimate the flux densities of the sources in the millimeter and  
sub-millimeter bands. We extrapolated and summed the ionized-gas  
(free-free radio emission) and dust (thermal emission) contributions in the afore mentioned wavelength ranges.
The possibility of Planck detecting the selected sources can be assessed 
by comparing the estimated flux densities to the expected sensitivity in each Planck channel. 
To obtain a realistic estimation of the noise produced by the Galactic emission, 
the Planck sky model software package was used.
}
%results heading (mandatory)
{
For each target source, from our new 7~mm data and other radio measurements from the literature,  
important physical parameters such as electron density and their spatial distribution, 
source geometry and emission measure were derived.
We conclude that, in the case of the present sample, located close to the Galactic center,
Planck will have a very low detection rate.
In contrast, assuming that our sample is representative of the whole UCHII-region population, 
we derive a very high probability of detecting this kind of source with Planck if located instead close to the anticenter.
From the analysis of the ionized-gas properties, we suggest that the selected sample
could also be contaminated by other kinds of Galactic objects.
}
% conclusions heading (optional)
{} %leave it empty if necessary


     \maketitle
%
%________________________________________________________________

\section{Introduction}

The Planck mission of the European Space Agency (ESA) will operate 
over a wide range of frequencies, from 30 to 900 GHz (from 1 cm to about 300 $\mu$m).
The frequency coverage of the Planck channels was selected to correspond to the so-called
cosmic microwave background (CMB) window, 
where temperature fluctuations due to Galactic emission are minimal.
Although Planck was designed  for cosmological 
studies, its multi-frequency nearly-full-sky maps
will have a profound impact on fundamental physics
as well as Galactic and extragalactic astrophysics 
and several Galactic projects are indeed included in its core program
(see \citet{Planck_col05} for details about the scientific case for the Planck
mission).

%%%%%%%%%%%%%%%%%%%%%%%%%%%%%%%%%%%%%%%%%%%%%%%%%%%%%%%%%%%%%%%%%%%%%%%%%%%%%%
\begin{table*}
%\caption{Observations at 7 mm. Flux density measurements ($S_\mathrm{7}$) with related errors ($\sigma_\mathrm{7}$)
%are reported.}
\caption{Observations.}
\label{tab}
\begin{center}

\vspace{0.3cm}

\begin{tabular}{lllrr|lllrr}
\hline
\hline
\s MSX6C Name   &\s R.A. (J2000) &\s Dec. (J2000) &\s $S_\mathrm{7}$ &\s $\sigma_\mathrm{7}$   &\s MSX6C Name  &\s R.A. (J2000)  
&\s Dec. (J2000) &\s $S_\mathrm{7}$ &\s $\sigma_\mathrm{7}$   \\
&\s [\ h\ m\ s] &~~\s [$^{\circ}\ ^{\prime}\ ^{\prime\prime}$] &\s [mJy]&\s [mJy]   & &\s [\ h\
m\ s] &~~\s [$^{\circ}\ ^{\prime}\ ^{\prime\prime}$] &\s [mJy]&\s [mJy]   \\
\hline                                                                
\s$17.2231+0.3952$        &\s18 20 46.13 &\s$-13$ 45 21.4 &\s$595$  &\s$90$  &\s$29.2113-0.0690$        &\s18 44 53.36 &\s$-03$ 20 28.1 &\s$150$  &\s$50$  \\
\s$16.9440-0.0736^{\dag}$ &\s18 21 56.0  &\s$-14$ 13 28   &\s$770$  &\s$160$ &\s$29.9564-0.0174$        &\s18 46 04.07 &\s$-02$ 39 16.6 &\s$3320$ &\s$300$ \\
\s$18.4608-0.0034$        &\s18 24 36.34 &\s$-12$ 51 00.0 &\s$330$  &\s$70$  &\s$30.5345+0.0209$        &\s18 46 59.38 &\s$-02$ 07 20.7 &\s$1070$ &\s$100$ \\
\s$18.7104+0.0001$        &\s18 25 04.15 &\s$-12$ 37 40.1 &\s$270$  &\s$70$  &\s$30.2340-0.1392$        &\s18 47 00.40 &\s$-02$ 27 47.9 &\s$250$  &\s$60$  \\
\s$18.3029-0.3910^{\dag}$ &\s18 25 42.5  &\s$-13$ 10 20   &\s$1260$ &\s$210$ &\s$30.8667+0.1141$        &\s18 47 15.71 &\s$-01$ 47 05.4 &\s$560$  &\s$110$ \\
\s$18.7084-0.1263$        &\s18 25 31.45 &\s$-12$ 41 18.5 &\s$240$  &\s$60$  &\s$31.4134+0.3092$        &\s18 47 34.35 &\s$-01$ 12 40.2 &\s$700$  &\s$120$ \\
\s$19.7403+0.2799$        &\s18 26 01.33 &\s$-11$ 35 08.4 &\s$160$  &\s$40$  &\s$30.6877-0.0729$        &\s18 47 36.01 &\s$-02$ 01 44.4 &\s$880$  &\s$140$ \\
\s$18.8330-0.3004$        &\s18 26 23.64 &\s$-12$ 39 35.9 &\s$280$  &\s$60$  &\s$30.7206-0.0826$        &\s18 47 41.79 &\s$-02$ 00 18.6 &\s$570$  &\s$130$ \\
\s$21.8732+0.0076^{\dag}$ &\s18 31 02.4  &\s$-09$ 49 33   &\s$560$  &\s$100$ &\s$31.0702+0.0498$        &\s18 47 51.74 &\s$-01$ 38 00.4 &\s$220$  &\s$60$  \\
\s$21.3853-0.2537$        &\s18 31 03.90 &\s$-10$ 22 40.7 &\s$215$  &\s$50$  &\s$31.2803+0.0615$        &\s18 48 11.89 &\s$-01$ 26 26.0 &\s$310$  &\s$60$  \\
\s$22.7253-0.0106$        &\s18 32 41.91 &\s$-09$ 04 40.8 &\s$195$  &\s$50$  &\s$30.6671-0.3316$        &\s18 48 29.24 &\s$-02$ 09 57.0 &\s$205$  &\s$50$  \\
\s$23.9548+0.1494^{\dag}$ &\s18 34 25.3  &\s$-07$ 54 50   &\s$1170$ &\s$160$ &\s$31.2430-0.1108^{\dag}$ &\s18 48 45.1  &\s$-01$ 33 13   &\s$660$  &\s$110$ \\
\s$24.5076-0.2222^{\dag}$ &\s18 36 46.9  &\s$-07$ 35 38   &\s$630$  &\s$100$ &\s$32.1514+0.1317^{\dag}$ &\s18 49 32.7  &\s$-00$ 38 04   &\s$680$  &\s$100$ \\
\s$25.5196+0.2156$        &\s18 37 05.20 &\s$-06$ 29 32.5 &\s$360$  &\s$90$  &\s$31.3948-0.2585$        &\s18 49 33.28 &\s$-01$ 29 07.5 &\s$380$  &\s$90$  \\
\s$25.2666-0.1584$        &\s18 37 57.97 &\s$-06$ 53 25.9 &\s$220$  &\s$50$  &\s$32.4727+0.2041$        &\s18 49 52.56 &\s$-00$ 18 53.1 &\s$220$  &\s$50$  \\
\s$25.3977-0.1406^{\dag}$ &\s18 38 08.2  &\s$-06$ 45 57   &\s$2540$ &\s$170$ &\s$32.7977+0.1903$        &\s18 50 30.86 &\s$-00$ 01 52.7 &\s$3800$ &\s$300$ \\
\s$25.3820-0.1813$        &\s18 38 15.15 &\s$-06$ 47 53.8 &\s$2730$ &\s$300$ &\s$32.2718-0.2260$        &\s18 51 02.37 &\s$-00$ 41 19.9 &\s$280$  &\s$70$  \\
\s$26.0905-0.0578^{\dag}$ &\s18 39 07.2  &\s$-06$ 06 44   &\s$165$  &\s$50$  &\s$33.8104-0.1869^{\dag}$ &\s18 53 42.4  &\s$+00$ 41 48   &\s$400$  &\s$80$  \\
\s$26.6085-0.2122$        &\s18 40 37.47 &\s$-05$ 43 15.4 &\s$165$  &\s$50$  &\s$35.4672+0.1381$        &\s18 55 34.11 &\s$+02$ 19 15.7 &\s$350$  &\s$60$  \\
\s$27.4938+0.1893^{\dag}$ &\s18 40 49.2  &\s$-04$ 45 06   &\s$500$  &\s$60$  &\s$35.5736+0.0678$        &\s18 56 00.96 &\s$+02$ 22 59.2 &\s$300$  &\s$60$  \\
\s$27.9777+0.0780$        &\s18 42 06.41 &\s$-04$ 22 14.5 &\s$220$  &\s$70$  &\s$35.5789-0.0304$        &\s18 56 22.56 &\s$+02$ 20 32.7 &\s$540$  &\s$90$  \\
\s$28.2461+0.0134$        &\s18 42 49.66 &\s$-04$ 09 45.7 &\s$185$  &\s$50$  &\s$38.5497+0.1636^{\dag}$ &\s19 01 07.6  &\s$+05$ 04 23   &\s$165$  &\s$40$  \\
\s$28.2007-0.0494$        &\s18 42 58.12 &\s$-04$ 13 55.7 &\s$980$  &\s$160$ &\s$38.8750+0.3088$        &\s19 01 12.49 &\s$+05$ 25 48.1 &\s$390$  &\s$60$  \\
\s$28.6096+0.0170^{\dag}$ &\s18 43 28.9  &\s$-03$ 50 18   &\s$420$  &\s$90$  &\s$39.8821-0.3457$        &\s19 05 24.11 &\s$+06$ 01 31.4 &\s$300$  &\s$60$  \\
\s$28.6520+0.0271$        &\s18 43 31.40 &\s$-03$ 47 40.6 &\s$300$  &\s$70$  &\s$41.7410+0.0972^{\dag}$ &\s19 07 15.5  &\s$+07$ 52 42   &\s$310$  &\s$40$  \\
\s$28.2875-0.3639$        &\s18 44 15.09 &\s$-04$ 17 51.3 &\s$585$  &\s$120$ &                          &              &                &         &        \\
\hline      
\end{tabular}
\begin{list}{}{}
\item[$^{\dag}$] Reported as a multi source in the 6 cm catalog but associated with the same MSX source. 
The 7~mm observations were carried out by pointing at the MSX coordinates.
\end{list}

\end{center}
\end{table*}
%%%%%%%%%%%%%%%%%%%%%%%%%%%%%%%%%%%%%%%%%%%%%%%%%%%%%%%%%%%%%%%%%%%%%%%%%%%%%%

In the science of compact Galactic objects,
Planck will be sensitive to the millimeter emission from dusty envelopes of 
young and old stars and 
to the thermal radio emission from the ionized gas surrounding many classes of stellar objects.

HII regions of size about 0.1 pc, emission measure higher than 10$^7$ pc cm$^{-6}$,
electron density higher than $10^{4}$ cm$^{-3}$, 
and ionized mass of around $10^{-2}$~M$_{\odot}$ are classified as
ultra compact HII regions (UCHIIs) and are sites of high-mass star formation. 
In these regions, newly born stars are still embedded in their natal molecular clouds.
The ambient dust thermally reradiates the entire luminosity of the central star in the infrared (IR), 
making  UCHIIs among the brightest IR sources in the Galaxy.
The radiation fields of the massive hot stars inside UCHIIs (OB spectral type) 
ionize the circumstellar material. The
signature of ionization is the typical free-free flux-density distribution at radio wavelengths
($S_{\nu} \propto \nu^{\alpha}$). 
Therefore, we expect the Galactic population of UCHIIs
to be a strong foreground contamination source to the Planck experiment.

As part of the preparation of a Planck pre-launch catalog,
we started an observing program to measure the 7~mm flux density of 
samples of Galactic sources with the 32~m INAF-IRA Noto Radiotelescope.
In a previous paper \citep{umana_etal08},
we published the our flux density measurements of planetary nebulae (PNe).

In this paper, we present new 7~mm single-dish observations of UCHIIs.
Main goal was to obtain reliable estimates of the flux densities expected in Planck channels
by building and modeling the spectral energy distribution (SED) of our targets. 
The observing frequency (43 GHz) is within the cosmic microwave background (CMB) window and
represents one of the channels of the forthcoming Planck mission.
Therefore, in this frequency band, we obtain a direct measurement of 
the expected flux density with minimal frequency extrapolation.

As added value, the present work provides a sizeable dataset of 7~mm measurements 
of UCHII regions. These data provide strong constraints on the observed SEDs 
in the spectral region where free-free emission dominates.

The SED modeling in the radio-millimeter spectral range is a crucial step 
for the study of the physics of ionized envelopes around UCHII regions.
The 7~mm measurements presented in this paper provide the possibility of constraining 
parameters such as the electron density and the radial profile, the
source geometry, and the emission measure.

\section{Sample selection}
\label{s_s}
Our sample was selected from the catalog of \citet{giveon_etal05}, where a cross-correlation between
the Midcourse Space Experiment (MSX6C) catalog \citep{egan_etal03}
and the Multi-Array Galactic Plane Imaging Survey (MAGPIS) \citep{white_etal05} was performed. 
MAGPIS was carried out at 6 cm
and covered Galactic regions with longitude $350^{\circ} \leq l \leq 42^{\circ}$
and latitude $|b| \leq 0.4^{\circ}$.
A comparison of the two catalogs (radio and infrared) allowed the authors to find a sample of 687 radio
sources with highly-reliable infrared counterparts. 
Only 228 of 687 targets were detected in the new MAGPIS at 20~cm \citep{helfand_etal06}.

All  sources with both 6 and 20~cm data
have red MSX colors and flat or inverted spectral indices, as is typical of thermal radio sources.
The average spectral index between 20 and 6~cm is equal to 0.28.
\citet{giveon_etal05} classified these sources as compact 
HII regions on the basis of their radio and infrared properties. For most sources, this was their first classification.
Starting from this subsample of 228, we first inspected all the corresponding MAGPIS 6~cm images.

To minimize the source-structure effect on the flux density measurements,
we selected only sources 
compact enough to be considered point-like at the resolution of the Noto Radiotelescope at 7~mm (HPBW about $1^{\prime}$) and rejected
those located in high-confusion regions.
In our sample, we included single sources of 6~cm angular size $\leq 20^{\prime\prime}$
and sources that, although mapped as multiple at 6~cm, showed separations $\leq 10^{\prime\prime}$ 
and were associated with only one infrared counterpart.
Finally, from the resulting sample we selected only sources
with integrated flux densities higher than 100~mJy at 6~cm and declination higher than $-15^{\circ}$.
In the conservative hypothesis of an optically thin nebula at 6~cm, a cut-off at both 100~mJy and $\delta>-15^{\circ}$
guarantees  7~mm detectability and observability with the Noto Radiotelescope.
The expected 7~mm flux density is a lower limit
in the case of an optically thick nebula at 6~cm.
All the above selection criteria reduced our sample to a total 
of 51 UCHIIs.

The names and positions of the selected targets are reported 
in Cols. 1, 2, and 3 and 6, 7, and 8 of Table~\ref{tab}. 
When observing multiple sources associated with the same infrared counterpart, we pointed the telescope at the MSX coordinates.

\section{Observations}

The observations reported in this paper were carried out at different epochs 
between January 2007  and July 2008, using the 32~m Radiotelescope
in Noto (Italy). The telescope has an active surface system that
accounts for the gravitational deformation of the primary mirror,
providing high quality observations at high frequencies (see \citet{orfei_etal02, orfei_etal04}
for details of the active surface system mounted on the Noto radiotelescope).
The 7~mm receiver used in our observations is a cooled
superheterodyne receiver
of typical zenith system temperature ($T_\mathrm{sys}$) about 100~K.
The observations were performed with a 400~MHz instantaneous band
centered on the sky frequency $\nu _\mathrm{sky} =42950$~MHz, the beam size
being $0.9^{\prime}$.
The gain ranges from 0.04 to 0.08~K/Jy, depending on the air mass as a function of the elevation.
Our targets were observed using the On The Fly (OTF)
scan technique, which consists of driving the beam
of the telescope across the sources in the RA direction.
The typical scan duration was approximately 20~s, which is
short enough to remain close to the white-noise
regime of the radiometer. The scan length was $3^{\prime}$, corresponding
to about three telescope beams.
To achieve a good signal-to-noise ratio,
each source was observed several times, for a total
integration time of 30 minutes.
Multiple OTF scans were then added together.

Daily gain curves were obtained and the flux
scale fixed using NGC~7027 as a primary
calibrator and 3C286 and DR21 as secondary calibrators. 
The adopted flux density for NGC~7027 is 5.07 Jy.
This value was obtained by correcting that reported in \citet{ott_etal94}
for the observed evolution of the spectral flux density in NGC~7027 \citep{zijlstra_etal08}.

All measurements were corrected for atmospheric extinction. 
The correction factor was calculated by performing a series of tipping scans
at elevations between $20^{\circ}$ and $80^{\circ}$,
covering a range of $\sec (z)$ from 1.01 to 3. Tipping scans were performed several times (once per hour) during each observing run.
The opacity correction was applied to each flux measurement after interpolating 
the results of contiguous tipping scans.
The sky-opacity  measurement averaged over all of the observing sessions was about $0.14$. 
The opacity fluctuation ranges from about 7\% (measured during observing sessions
with very stable sky conditions) to about 30\% (observing sessions
characterized by worse weather).

%%%%%%%%%%%%%%%%%%%%%%%%%%%%%%%%%%%%%%%%%%%%%%%%%%%%%%%%%%%%%%%%%%%%%%%%%%%%%%%%%%%%%%%%%%%%%%%%%%%%%%%%%%%%%%%%%%%%%%%%%%%%%%%%%%%%%%%%%%%
\begin{table*}
\caption{Sources parameters.}
\label{tab2}
\begin{center}
\vspace{0.3cm}
\begin{tabular}{lrc|lrc|lrc}
\hline          
\hline
\s MSX6C Name &\s $\alpha$ &\s $EM$                    &\s MSX6C Name &\s $\alpha$ &\s $EM$                    &\s MSX6C Name &\s $\alpha$ &\s $EM$                    \\
        &      &\s $\times 10 ^7$ [pc cm$^{-6}$] &        &      &\s $\times 10 ^7$ [pc cm$^{-6}$] &        &      &\s $\times 10 ^7$ [pc cm$^{-6}$] \\
\hline
\s$17.2231+0.3952^{\dag}$ &\s$0.54$  &\s$>3.12~~~~$   &\s$26.0905-0.0578$        &\s$0.11$  &\s$0.11$   &\s$31.0702+0.0498$        &\s$0.34$  &\s$0.19$  \\
\s$16.9440-0.0736$        &\s$0.23$  &\s$6.17 $   &\s$26.6085-0.2122$        &\s$-0.09$ &\s$0.26$   &\s$31.2803+0.0615$        &\s$0.01$  &\s$0.21$  \\
\s$18.4608-0.0034$        &\s$0.02$  &\s$3.39 $   &\s$27.4938+0.1893$        &\s$0.31$  &\s$0.03$   &\s$30.6671-0.3316$        &\s$0.13$  &\s$4.77$  \\
\s$18.7104+0.0001^{\dag}$ &\s$0.45$  &\s$>12.80~~~~~~$  &\s$27.9777+0.0780$        &\s$0.27$  &\s$0.11$   &\s$31.2430-0.1108$        &\s$0.16$  &\s$1.12$  \\
\s$18.3029-0.3910$        &\s$-0.13$ &\s$0.11 $   &\s$28.2461+0.0134$        &\s$0.24$  &\s$0.10$   &\s$32.1514+0.1317$        &\s$-0.14$ &\s$0.06$  \\
\s$18.7084-0.1263$        &\s$-0.03$ &\s$3.76 $   &\s$28.2007-0.0494^{\dag}$ &\s$0.58$  &\s$>2.21~~~~$  &\s$31.3948-0.2585$        &\s$0.53$  &\s$0.13$  \\
\s$19.7403+0.2799$        &\s$0.07$  &\s$0.03 $   &\s$28.6096+0.0170$        &\s$0.21$  &\s$3.25$   &\s$32.4727+0.2041$        &\s$0.34$  &\s$3.12$  \\
\s$18.8330-0.3004$        &\s$0.41$  &\s$0.50 $   &\s$28.6520+0.0271$        &\s$0.09$  &\s$0.61$   &\s$32.7977+0.1903^{\dag}$ &\s$0.31$  &\s$>3.41~~~~$ \\
\s$21.8732+0.0076$        &\s$-0.03$ &\s$0.13 $   &\s$28.2875-0.3639$        &\s$-0.06$ &\s$1.89$   &\s$32.2718-0.2260$        &\s$-0.15$ &\s$0.25$  \\
\s$21.3853-0.2537^{\dag}$ &\s$0.32$  &\s$>2.49~~~~$   &\s$29.2113-0.0690$        &\s$0.08$  &\s$0.73$   &\s$33.8104-0.1869$        &\s$0.32$  &\s$0.03$  \\
\s$22.7253-0.0106$        &\s$0.28$  &\s$0.07 $   &\s$29.9564-0.0174$        &\s$0.09$  &\s$3.00$   &\s$35.4672+0.1381$        &\s$0.04$  &\s$0.76$  \\
\s$23.9548+0.1494$        &\s$-0.10$ &\s$0.12 $   &\s$30.5345+0.0209$        &\s$0.08$  &\s$1.24$   &\s$35.5736+0.0678$        &\s$0.19$  &\s$0.26$  \\
\s$24.5076-0.2222^{\dag}$ &\s$0.35$  &\s$>0.76~~~~$   &\s$30.2340-0.1392$        &\s$-0.15$ &\s$3.16$   &\s$35.5789-0.0304^{\dag}$ &\s$0.62$  &\s$>8.96~~~~$  \\
\s$25.5196+0.2156$        &\s$0.04$  &\s$0.29 $   &\s$30.8667+0.1141$        &\s$0.34$  &\s$5.38$   &\s$38.5497+0.1636$        &\s$0.03$  &\s$0.03$  \\
\s$25.2666-0.1584$        &\s$-0.09$ &\s$10.71~$   &\s$31.4134+0.3092$        &\s$-0.13$ &\s$0.56$   &\s$38.8750+0.3088$        &\s$0.08$  &\s$2.25$  \\
\s$25.3977-0.1406$        &\s$0.00$  &\s$0.60 $   &\s$30.6877-0.0729$        &\s$0.06$  &\s$1.72$   &\s$39.8821-0.3457$        &\s$-0.07$ &\s$1.75$  \\
\s$25.3820-0.1813$        &\s$0.29$  &\s$0.80 $   &\s$30.7206-0.0826$        &\s$0.11$  &\s$3.76$   &\s$41.7410+0.0972$        &\s$-0.09$ &\s$0.08$  \\
\hline                            
\end{tabular}                                                         
\begin{list}{}{}
%\item[$^{\ast}$]{Spectral indices ($\alpha$) evaluated between 6~cm and 7~mm.} 
%\item[$^{\ast\ast}$]{Emission measures ($EM$) derived from Eq.~\ref{em}.} 
\item[$^{\dag}$] {For these sources the simulated radio spectra indicate that the emission is optically thick up to 7~mm,
the derived $EM$ has to be considered as a lower limit.}
\end{list}

\end{center}                                                          
\end{table*}                                                          
%%%%%%%%%%%%%%%%%%%%%%%%%%%%%%%%%%%%%%%%%%%%%%%%%%%%%%%%%%%%%%%%%%%%%%%%%%%%%%%%%%%%%%%%%%%%%%%%%%%%%%%%%%%%%%%%%%%%%%%%%%%%%%%%%%%%%%%%%%%

The telescope pointing was checked before each observing session
by measuring its accuracy for a large sample of pointing calibrators:
SiO maser sources and bright sources at 7 mm.
The estimated pointing accuracy is about $10^{\prime\prime}$.

Taking into account the telescope pointing accuracy, the error in the flux density of the primary
gain calibrator, the errors in the opacity measurements,
and the gain-curve error, 
we estimate the overall calibration accuracy to be around 10--15$\%$.

We detected all targets with a 3$\sigma$ confidence threshold, with a detection rate of 100$\%$. 
The results of these observations are reported in Table~\ref{tab}, 
where the measured 7~mm flux densities ($S_\mathrm{7}$)  are listed with their errors (Cols. 4, 5 and 9, 10).

\section{The SEDs}

\subsection{The dust contribution}
\label{d_c}

%%%%%%%%%%%%%%%%%%%%%%%%%Figural 5%%%%%%%%%%%%%%%%%%%%%%%%%%%%%%%%%%%%%%
\begin{figure}
\resizebox{\hsize}{!}{\includegraphics{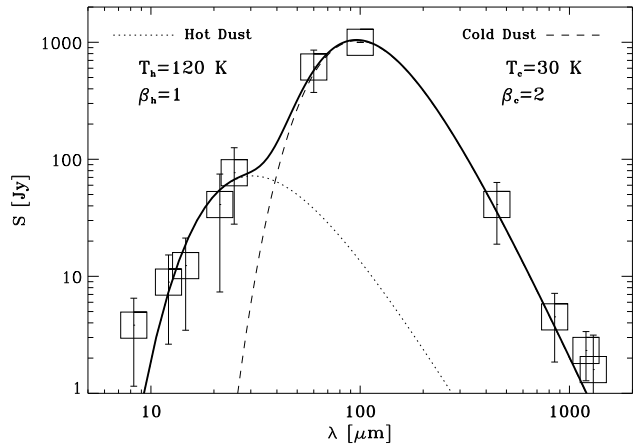}}
\caption{SED template of the dust contribution.
The data collected from the far-IR to the mm range have been normalized to 1000 Jy at 100 $\mu$m.
The squares represent the mean values of the normalized measurements and
standard deviations are also shown. 
Two dust components are evident: the hot component determines the contribution
at shorter wavelengths (dotted line), while the cold component dominates in the submillimeter
spectral range (dashed-line).}
\label{sed_media}
\end{figure}
%%%%%%%%%%%%%%%%%%%%%%%%%Figura%%%%%%%%%%%%%%%%%%%%%%%%%%%%%%%%%%%%%%

To determine the dust contribution for our sample of UCHIIs, 
we firstly searched both the literature and online catalogs for
data in the range from 8~$\mu$m to about 1~mm,
where the dust contribution is relevant.
As a consequence of the selection criteria, MSX data \citep{egan_etal03} are
available for all of the 51 sources of the sample.
We found IRAS data \citep{helou_&_walker88}
for 45 sources, Scuba data for
12 sources at 450~$\mu$m 
and for 16 at 850~$\mu$m 
\citep{williams_etal04, wu_etal05, thompson_etal06},
and  Calthech Submillimeter Observatory (CSO) data  for
4 sources at 350~$\mu$m \citep{hunter_etal00, motte_etal03}.
For 8 sources, we found measurements at 1.2~mm 
carried out with the IRAM~30~m Radiotelescope 
\citep{altenhoff_etal94, beuther_etal02} 
and SIMBA \citep{faundez_etal04}.
Ten sources were also observed at 1.3~mm at the 
IRTF on Mauna Kea \citep{chini_etal86a, chini_etal86b},
at the SEST \citep{launhardt_&_henning97}, 
and with the MAMBO bolometer array installed on the IRAM~30~m Radiotelescope \citep{motte_etal03}.
To summarize, we have enough measurements to compile SEDs for 34 sources in our sample.
From the SEDs, we derive an SED template 
by normalizing  the 100~$\mu$m high-quality IRAS data to 1000~Jy and then averaging together
all the individual SEDs.

The SED template obtained can be modeled by using two dust components at different temperatures. 
We assume that each of the two dust components 
emits as a blackbody modified by the frequency-dependent dust opacity. 
This gives $S_{\nu}\propto \nu^\beta B_{\nu}(T_\mathrm{d})$,
where $B_\nu(T_\mathrm{d})$ is the Planck function for a dust temperature of $T_\mathrm{d}$
and where $\beta $ is the emissivity index. The index $\beta $ strongly depends on the mineralogical composition
of the grains and on their physical shapes.
Following the results in \citet{chini_&_krugel86} about modeling dust emission from
UCHII regions, we find that for the cold dust component a good fit is achieved 
for a dust temperature of 30~K and an emissivity index $\beta=2$.
The hot component has been fitted for a dust temperature of 120~K
and by assuming that $\beta= 1$ \citep{faundez_etal04}.

In Fig.~\ref{sed_media}, we show the averaged data and overplot their best-fit solution.
The deviation of the point at 8~$\mu$m could be related to the contribution
of either a very hot dust component or
the stellar component, whereas the deviation of the two measurements
at 1.2 and 1.3 mm can be explained as evidence of an ionized gas contribution. 

For 17 sources, we do not have enough measurements
at $\lambda \geq 100$ $\mu$m to compile their individual SEDs.
To estimate their SED contribution from cold dust, we assume that all sources in our sample have the same dust properties. 
Our SED template is therefore a good assessment of their emission curves 
(Fig.~\ref{sed_media}) and we can obtain an individual SED for each of them by just scaling the template to their MSX 21 $\mu$m measurement.
In the SED template, the flux density at 21 $\mu$m is found to be a factor of 0.04 of its value at 100 $\mu$m.

\subsection{The ionized-gas contribution}
\label{f_c}

We estimate the contribution from ionized gas 
to the SEDs of the selected sources
by combining our 7~mm measurements with the flux densities at 6~cm in \citet{giveon_etal05} and those at 20~cm, 
the observations at the lower frequency being available for each source in our sample 
from the new MAGPIS catalog reported by \citet{giveon_etal05b}.

In the present work, we need to assess the contribution from free-free emission in an angular element of size
comparable to the Noto Radiotelescope beam (HPBW about $1^{\prime}$). For those sources with multiple peaks in their 6~cm and 20~cm
high-resolution observations, we use as an estimation of their 6~cm and 20~cm 
flux densities the sum of the integrated fluxes of each peak.
Literature data at radio wavelengths shorter than 6~cm
were also collected for 18 targets in our sample 
\citep{wink_etal82, wood_etal88a, wood_etal88b, kurtz_etal94, gehrz_etal95, 
fey_etal95, walsh_etal98, martin_etal03, longmore_etal07}.

The spectral index $\alpha$ between 6~cm \citep{giveon_etal05} and 7~mm (this paper)
was calculated for each source.
The values obtained are reported in Table~\ref{tab2}.
The free-free emission from the ionized gas was derived by modeling all the available 
radio measurements for every source.

As described by \citet{hoare_etal91},
a typical UCHII region can be modeled
using spherical symmetry.
The source can be modeled as a central cavity surrounded by a shell
of ionized gas, the newly born high-mass star being located inside the cavity center.
The source geometry is characterized by the linear diameter and the ratio
of the internal to external radii ($r_\mathrm{int}$/$r_\mathrm{ext}$).
The shell of ionized material is located in the inner region of the dust cloud
responsible for the  infrared emission.
In this simple spherical-geometry model, the spatial density structure 
of the ionized gas is described by the power law $n_{\mathrm e}(r) \propto r^{-\gamma}$.

Adopting the model software developed by \citet{umana_etal08b},
the free-free radio spectrum was simulated for each source by assuming a fixed
value of the source linear diameter. 
In the following, we consider two cases of the 
source linear diameter (equal to $2\times r_\mathrm{ext}$): 0.1 pc and 0.05 pc.
The free-free model spectrum was calculated for different values of the ratio $r_\mathrm{int}$/$r_\mathrm{ext}$,
the index of the power law that describes the ionized-gas spatial distribution, and 
the corresponding average electron density. 

The distance to the source was treated as a free parameter in obtaining the best-fit 
SED to both the observations and simulations.
The source angular size
derived from the assumed linear diameter and distance needs to be comparable to 
the angular size measured at 6~cm \citep{giveon_etal05}.

\begin{figure}
\resizebox{\hsize}{!}{\includegraphics{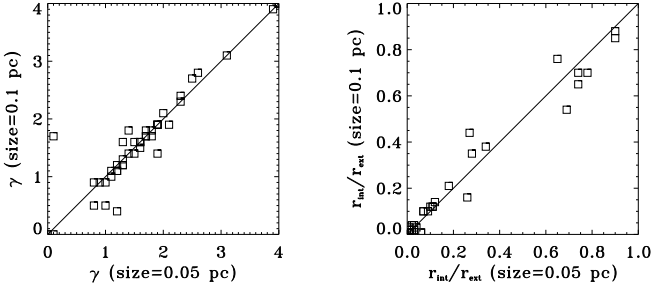}}
\caption{Comparison between the derived model parameters
for the two analyzed source diameters (0.1 pc and 0.05 pc). 
\textit{Left panel}: index of the power law that describes the density gradient
inside the source. \textit{Right panel}: ratio of internal to external source radii.}
\label{gamma_rr}
\end{figure}

The new 7~mm measurements allow us to characterize well
the free-free radio spectrum of each source in our sample.
The distance-independent parameters are, in fact, univocally defined. 
The values of $\gamma$ and $r_\mathrm{int}$/$r_\mathrm{ext}$
for the two sizes do not
show any significant difference, as shown in Fig.~\ref{gamma_rr}.
Using our model we found that
the index $\gamma$ covers wide range of model solutions,
many sources being reproduced well
by assuming homogeneous electron-density radial distribution,
while others have distributions that are highly inhomogeneous. 
In all cases, the geometry of most of the sources could be described
by the ratio $r_\mathrm{int}$/$r_\mathrm{ext}$ below 0.2.
The derived electron density $n_{\mathrm e}$ is on the order of $10^4$ cm$^{-3}$.

Following \citet{terzian_&_dickey73}, it is possible to derive the mean emission measure ($EM$)
if the observed radio optically thin 
flux density ($S_7$, in mJy) and the corresponding
source angular size ($\theta$, in arcsec) are known,

\begin{equation}
EM =\frac{5.3 \times 10^5 S_7}{\theta ^2}~~~~~~~~~~~~~~\mathrm{[pc~cm^{-6}].}
\label{em}
\end{equation}

\noindent
To derive the mean emission measure from our 7~mm measurements,
we assume that the source angular size at 7~mm equals the angular size at 6~cm measured 
by \citet{giveon_etal05}. The derived $EM$ is reported in Table~\ref{tab2}.
For sources with simulated spectra that are optically thick up to 7~mm,
the derived $EM$ is considered to be a lower limit.

Note that the spectral index between 6~cm and 7~mm 
by itself is not able to characterize the emission at 7~mm.
In many cases, sources
with positive spectral index are characterized by optically thin emission at 7~mm.

\begin{figure}
\resizebox{\hsize}{!}{\includegraphics{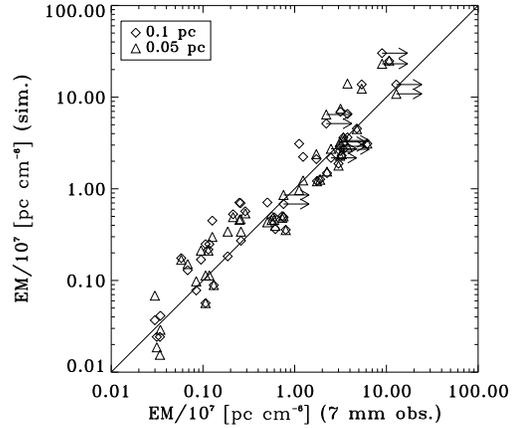}}
\caption{Comparison of the Emission Measure deduced
from the simulation with those calculated from the 7~mm measurements.
Arrows refer to those sources optically thick ut to 7~mm.}
\label{simulaz}
\end{figure}

Our modeling also allows us to estimate the 
emission measure. In Fig.~\ref{simulaz},
the emission measure obtained from the model fit
is plotted as a function
of the $EM$ calculated from Eq.~\ref{em}.

The model used to simulate the radio emission from UCHII regions assumes 
the highly simplified hypothesis of spherical geometry, whereas
radio images at high angular resolution show that these sources 
are characterized by various morphologies \citep{wood_&_churchwell89a,kurtz_etal94,
depree_etal05}, i.e., shell-like, bipolar, cometary, irregular, and spherical. 
In spite of the possible difference between the real and the modeled source geometry,
the clear correlation between the values of $EM$ deduced from the model and those
calculated from the observations indicates that our model parameters allow us to 
reproduce well the overall properties of the targets.

%%%%%%%%%%%%%%%%%%%%%%%%%Figural 5%%%%%%%%%%%%%%%%%%%%%%%%%%%%%%%%%%%%%%
%%%%%%%%%%%%%%%%%%%%%%%%%%%%%%%%%%%%%%%%%%%%%%%%%%%%%%%%%%%%%%%%%%%%%%%%%%%%%%

Our calculations exhibit a large spread in $EM$ (Fig.~\ref{simulaz}). HII regions characterized by $EM \geq 10^7$~pc~cm$^{-6}$
are defined as ultra compact.
Sources with $EM$ values
at the edges of the range do not show any 
significant difference in their dust emission compared to the other sources of the sample.
No relation was observed between infrared colors and $EM$.
This physical parameter is also randomly distributed 
as a function of the Galactic longitude.

The infrared colors of the selected sample follow a typical trend for UCHII regions, 
while in the radio-millimeter regime not all sources have typical UCHII characteristics,
as illustrated by their $EM$ values.
Our study in the radio range demonstrates that
a sample of UCHIIs selected only on the basis of their infrared colors
may be contaminated by other classes of Galactic sources. Planetary nebulae or high-mass evolved stars
could have infrared colors similar to those of UCHII, but different ionized gas properties.

\section{Estimated detectability with Planck}

Planck is equipped with a Low Frequency Instrument (LFI), 
operating at 30, 44, and  70 GHz and a High Frequency Instrument (HFI) operating at 100, 143, 217, 353, 545, and 857 GHz.
To evaluate the fraction of UCHIIs in our sample that is detectable with Planck,
we need to compare the modeled flux densities of the sources with the expected sensitivity in each observational band.

\begin{figure}
\resizebox{\hsize}{!}{\includegraphics{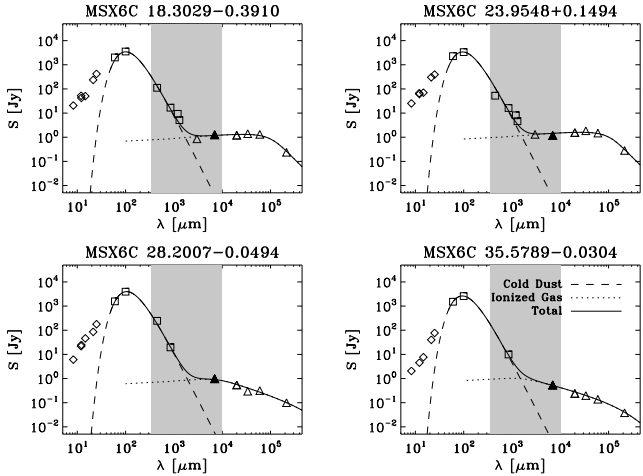}}
\caption{Examples of the source SEDs, different symbols refer to different components.
Diamonds: hot dust component; squares: cold dust component; triangles: ionized gas.
The dust component is deduced from the template SED, the free-free emission of the ionized gas derived
from its model. The spectral range covered by Planck is represented by the shaded area.}
\label{4sed}
\end{figure}

We derive the modeled flux densities at each Planck channel from the source SEDs 
by calculating the dust and gas emission as described in Sects.~\ref{d_c} and \ref{f_c} and then summing both contributions. 
In Fig.~\ref{4sed}, four of the derived  SEDs are shown as an example;
the shaded area represents the spectral range covered by the Planck satellite.
The dust contribution was obtained by extrapolating the modified-blackbody 
fit of the cold dust component (dashed lines in Fig.~\ref{4sed}), while the free-free contribution 
was obtained by modeling the data points as described in Sect.~\ref{f_c} and extending the
derived synthetic spectra to the submillimeter range (dotted line in Fig.~\ref{4sed}).

%%%%%%%%%%%%%%%%%%%%%%%%%Figural 5%%%%%%%%%%%%%%%%%%%%%%%%%%%%%%%%%%%%%%
%%%%%%%%%%%%%%%%%%%%%%%%%%%%%%%%%%%%%%%%%%%%%%%%%%%%%%%%%%%%%%%%%%%%%%%%%%%%%%

Following \citet{umana_etal06}, 
the expected total sensitivity is assumed  
to be the sum in quadrature of all the sources of confusion noise, namely (i)
the nominal instrumental Planck sensitivity per resolution element,
(ii) the extragalactic-foreground confusion noise \citep{toffolatti_etal98},
(iii) the CMB anisotropy \citep{bennett_etal03a}, and (iv) the Galactic-foreground confusion noise.

As pointed out by \citet{wood_&_churchwell89b},
the Galactic distribution of candidate massive-star formation sites 
is mainly confined to the Galactic plane, and our sample consists indeed of UCHIIs located in the Galactic plane, 
where we also expect the Galactic emission to be the major source of confusion noise in all Planck channels
because of the emission of the dust and gas located in the Galactic disk.

The Galactic noise reported by \citet{umana_etal06} was derived for  regions of the sky outside of
the Galactic plane, therefore we cannot use this value for UCHII regions.
For a realistic estimation of the noise related to the Galactic emission, we 
simulate the sky emission in the range between 30 and 900 GHz with
the Planck sky model (PSM) software package (version 1.6.3).
PSM is a set of programs and ancillary data developed by the Planck ESA Working Group 2 that can
simulate the sky emission at Planck frequencies and angular resolution, and reproduces the  
thermal free-free, synchrotron, and the dust components, i.e., thermal dust and spinning dust. %**********************************
\footnote{See \citet{delabrouille_etal09} and \citet{leach_etal08} for
applications and capabilities of the PSM package.}
The free-free component is modeled following \citet{dickinson_etal03} 
and the spatial structure estimated from the dust-extinction corrected $H_{\alpha}$ emission. 
The synchrotron  component is calculated by assuming a perfect power-law with different values of the spectral index  
in different sky directions, as derived from
the 408~MHz \citep{haslam_etal82} and WMAP 23~GHz \citep{bennett_etal03b} maps.
The thermal emission from interstellar dust
is assessed following \citet{finkbeiner_etal99}.
The prediction of the rotating grain emission is obtained 
by multiplying the column densities of the cold neutral medium 
\citep{schlegel_etal98} with the emissivities calculated using the spinning dust model developed by
\citet{draine_&_lazarian98}.

For the two Planck channels of lower frequency (30 and 44 GHz), the Galactic emission is dominated by the free-free component,
whereas for the third channel (70 GHz) the free-free and the thermal dust components are comparable;
for the other channels the emission of the thermal dust dominates.
The other emission mechanisms provide negligible large-scale contributions.

%%%%%%%%%%%%%%%%%%%%%%%%%%%Figural 5%%%%%%%%%%%%%%%%%%%%%%%%%%%%%%%%%%%%%%
%%%%%%%%%%%%%%%%%%%%%%%%%%%%%%%%%%%%%%%%%%%%%%%%%%%%%%%%%%%%%%%%%%%%%%%%%

\begin{figure}
\resizebox{\hsize}{!}{\includegraphics{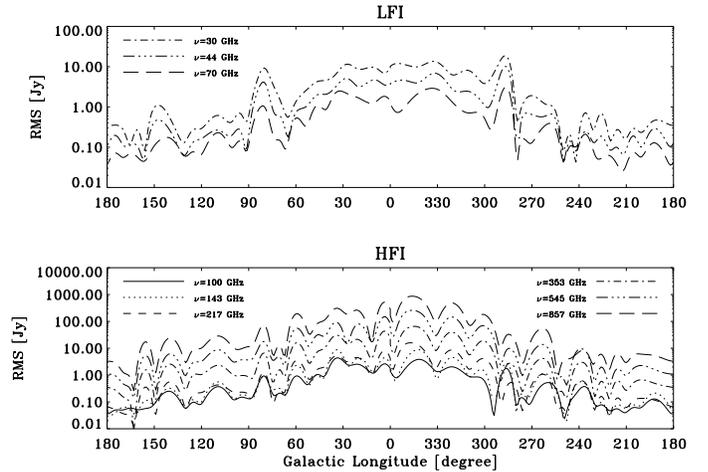}}
\caption{Expected Galactic emission fluctuation for $|b| \leq 0.5^{\circ}$ 
at different values of Galactic longitude for the two Planck instruments.}
\label{Galactic_noise}
\end{figure}

To obtain an idea of the trend of
the Galactic emission fluctuations as a function of the Galactic longitude, the
standard deviation of the Galactic emission for each PSM channel map was calculated in boxes
of size $3 \times \theta_{\nu}$ 
($\theta_{\nu}$ is the Planck beam at the considered frequency), 
centered on Galactic latitude $0^\circ$ and spaced every $5^{\circ}$ in longitude.
We selected the size of the region
in which the standard deviation of the Galactic emission was to be calculated as 
a compromise between the necessity both to have a large number of pixels in the region
and to remain well inside the plane of the Galaxy.
We show the RMS of the Galactic emission for each channel map at different values of Galactic longitude
for the two Planck instruments in Fig.~\ref{Galactic_noise}.
The Galactic emission fluctuation strongly depends on 
the longitude. The first and fourth Galactic quadrants (center) are characterized
by higher noise with respect to the second and third quadrants (anticenter),
as an obvious consequence of the larger amount of dust in the direction of the center
of the Galaxy.

The correctness of the PSM predictions was tested by comparing them to the observed 
all-sky WMAP temperature map for Stokes I at 41 GHz (Q band). 
The WMAP data were retrieved  from the NASA Legacy Archive for Microwave Background Data Analysis (LAMBDA),
while  a sky temperature map  at the same frequency and angular resolution was constructed with PSM. 
We found very good agreement between them, allowing us to use 
our results on the Galactic emission fluctuation with confidence.

\begin{figure}
\resizebox{\hsize}{!}{\includegraphics{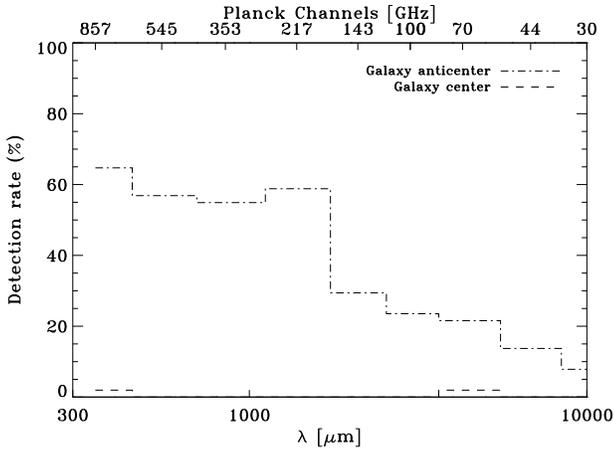}}
\caption{Expected Planck capability of detecting the sample of UCHII analyzed here.
\textit{Dashed line}: true source Galactic position. \textit{Dot dashed line}:
results for the same population but located in the anticenter region.
}
\label{sens_plank}
\end{figure}

In Fig.~\ref{sens_plank}, we show the percentage of objects from our sample that could be 
detected above the 3$\sigma$ threshold for each channel of the instruments onboard Planck.
Unfortunately, all the UCHIIs in our sample are located in the first and fourth quadrants
and because of the high Galactic emission fluctuation
a very small fraction of them should be detected by Planck.

However, UCHIIs have been detected in any Galactic direction \citep{wood_&_churchwell89b}.
Therefore, assuming that our sample is representative of the entire class of UCHIIs, 
we can also calculate the detectability 
of those located in the region of the Galactic anticenter.
Figure~\ref{sens_plank} shows how UCHIIs in the direction of the anticenter
should be easily seen by Planck.

Unfortunately, the high-mass star-formation rate is expected to strongly decrease far away from the center of the Galaxy.
\citet{bronfman_etal00} reported that the OB star-formation sites 
in the anticenter direction reduced to about 10$\%$ those detected in directions close to the center of the Galaxy. 
The expected increased Planck capability to detect UCHIIs in the anticenter direction has to
be compared with the decreasing star formation rate.
The true expected number of detectable UCHIIs must be weighted by the radial distribution of star formation in the Galaxy.

The UCHII sample analyzed in this paper represents only a subsample of the 228 UCHII regions 
reported by \citet{giveon_etal05}, which are distributed in a region of the Galactic disk 
that cover about $52^{\circ}$ of Galactic longitude.
Taking into account the estimation of \citet{bronfman_etal00}, the expected UCHIIs
that Planck could detect for a similar range of Galactic longitudes in the anticenter direction 
could be tens.
Note that the Planck channels that have the higher capability of detecting UCHIIs are the
channels at the lower wavelengths (Fig.~\ref{sens_plank}), where the dust emission contribution is dominant.
However, the sample selected by \citet{giveon_etal05} has a bias towards sources above the
threshold detectability at the radio wavelengths.
Therefore, it is reasonable to expect that the estimation reported here 
is only a lower limit to the true number of UCHII regions detectable by the Planck satellite,
as there may be many sources with very low radio emission levels, for actual sensitivities, 
but with detectable IR dust emission.
We emphasize that
this conclusion is supported by many detections of UCHIIs in the region of the Galactic anticenter obtained 
with the balloon experiment Archeops at mm and sub-mm wavelengths \citep{desert_etal08}.

%%%%%%%%%%%%%%%%%%%%%%%%%%%%%%%%%%%%%%%%%%%%%%%%%%%%%%%%%%%%%%%%%%%%%%%%%%%%%%
%%%%%%%%%%%%%%%%%%%%%%%%%%%%%%%%%%%%%%%%%%%%%%%%%%%%%%%%%%%%%%%%%%%%%%%%%%%%%%

In many cases UCHIIs have been observed on the borders of larger and fainter HII regions.
In higher wavelength channels, the Planck beam 
may be large enough to detect emission from these associated normal HII regions.
Considering their contribution, the Planck capability to detect UCHIIs here reported could be an underestimate.

We conclude that we expect the Planck mission to provide an important contribution   
to studies of UCHIIs, as it will provide the scientific community 
with measurements in a still largely-unexplored spectral region.

\section{Summary}

We have presented  7~mm (43~GHz) observations of a sample of radio-bright ultra-compact HII regions 
carried out with the INAF-IRA Noto Radiotelescope. 
The results of this survey provide a sizeable database of new millimeter measurements that allow us 
to characterize the high-frequency free-free contribution
from the ionized-gas components of the selected targets.
The free-free spectrum 
of each source were derived by modeling all the available radio and millimeter 
observations and the new 7~mm data.

We used our 7~mm measurements along with infrared and sub-millimeter data to compile SEDs 
from the radio to the far-IR and to estimate the expected flux densities in the 
spectral region between the cm and the sub-mm ranges.

%%%%%%%%%%%%%%%%%%%%%%%%%Figural 5%%%%%%%%%%%%%%%%%%%%%%%%%%%%%%%%%%%%%%
%%%%%%%%%%%%%%%%%%%%%%%%%%%%%%%%%%%%%%%%%%%%%%%%%%%%%%%%%%%%%%%%%%%%%%%

The study of properties of Galactic UCHIIs is one of the
secondary science objectives of the Planck core program \citep{Planck_col05}
therefore, it appears to be important
to test the capability of Planck to detect UCHII regions.
We, thus, estimated the fluctuation of the Galactic emission as a function of the Galactic longitude 
by using the Planck Sky Model software package.
In this paper, we note that we report for the first time 
a quantitative estimation of the emission fluctuation of the Galactic disk as seen by Planck.
As expected, the Galactic emission fluctuation is higher near
the center of the Galaxy.
When comparing the modeled radio/sub-mm flux densities
to the expected total sensitivity of the Planck satellite, 
we evaluate that few sources are
likely to be detected by Planck
in the direction of the Galactic center,
where the sources of our sample are located. 
If we assume that the sample  analyzed in the present paper is representative of the whole Galactic UCHII population,
we infer a higher rate of detectability by Planck  
in the direction of the Galactic anticenter,
where the  Galactic emission fluctuation is significantly lower.

Therefore, even though designed for cosmological studies, 
Planck should also contribute to the UCHII science. 
The rich source database provided by Planck will allow follow-up observations aimed at 
the investigating morphological details with future instruments working in the same spectral region, such as ALMA.

The source parameter emission measures 
were derived both directly from the 7~mm measurements and, independently, 
from a model of the ionized-gas continuum emission.
The derived $EM$ values for each sources are clearly correlated, supporting the goodness of the model fit.
The analysis of this parameter leads us to conclude that our 
sample could be contaminated by other types of Galactic sources, although all of the selected targets 
can be classified 
as UCHII regions because of their infrared colors.

%________________________________________________________________
\begin{acknowledgements}
Based on observations with the Noto Radiotelescope operated by INAF-Istituto di Radioastronomia.
This work has been partially supported by ASI through contract Planck LFI Activity of Phase E2.
The authors acknowledge the use of the Planck Sky Model, developed
by the Component Separation Working Group (WG2) of the Planck Collaboration.
We thank the referee for his/her constructive criticism which
enabled us to improve this paper.
\end{acknowledgements}

\bibliographystyle{aa}

\begin{thebibliography}{99}


\bibitem[Altenhoff et al., 1994]{altenhoff_etal94} 
	Altenhoff, W.J., Thum, C., \& Wendker, H.J., 1994, A\&A, 281, 161
\bibitem[Bennett et al., 2003a]{bennett_etal03a}
        Bennett, C.L., Halpern, M., Hinshaw, G., Jarosik, N., Kogut, A., Limon, M., 
	Meyer, S.S., Page, L., Spergel, D. N., Tucker, G.S., and 11 coauthors 2003a, ApJS, 148, 1
\bibitem[Bennett et al., 2003b]{bennett_etal03b}
	Bennett, C.L., Hill, R. S., Hinshaw, G., Nolta, M.R., 
	Odegard, N., Page, L., Spergel, D.N., Weiland, J.L., Wright, E.L., Halpern, M., and 6 coauthors
	2003b, ApJS, 148, 97
\bibitem[Beuther et al., 2002]{beuther_etal02}
	Beuther, H., Schilke, P., Menten, K.M., Motte, F., Sridharan, T.K., \& Wyrowski, F.,  2002, ApJ, 566, 945
\bibitem[Bronfman et al., 2000]{bronfman_etal00}
	Bronfman, L., Casassus, S., May, J., Nyman, L.-\AA., 2000, A\&A, 358, 521
\bibitem[Chini et al., 1986a]{chini_etal86a} 
	Chini, R., Kreysa, E., Mezger, P.G., \& Gemund, H.-P., 1986a, A\&A, 154, L8
\bibitem[Chini et al., 1986b]{chini_etal86b} 
	Chini, R., Kreysa, E., Mezger, P.G., \& Gemund, H.-P., 1986b, A\&A, 157, L1
\bibitem[Chini \& Kr{\"u}gel, 1986]{chini_&_krugel86} 
	Chini, R.  \& Kr{\"u}gel, E., 1986, A\&A, 167, 315
\bibitem[Delabrouille et al., 2009]{delabrouille_etal09}
	Delabrouille, J. et al. 2009, in preparation
\bibitem[Desert et al., 2008]{desert_etal08}
	Desert, F.-X., Macias-Perez, J.F., Mayet, F., Giardino, G., Renault, C.,
	Aumont, J., Benoit, A., Bernard, J.-Ph., Ponthieu, N., \& Tristram, M.,
	2008, A\&A, 481, 411
\bibitem[De Pree et al., 2005]{depree_etal05} 
	De Pree, C.G., Wilner, D.J., Deblasio, J., Mercer, A.J., \& Davis, L.E., 2005, ApJ, 624, L101
\bibitem[Dickinson et al., 2003]{dickinson_etal03}
	Dickinson, C., Davies, R.D., \& Davis, R.J., 2003, MNRAS, 341, 369
\bibitem[Draine \& Lazarian, 1998]{draine_&_lazarian98}
	Draine, B.T., Lazarian, A., 1998, ApJ, 508, 157
\bibitem[Egan et al., 2003]{egan_etal03} 
	Egan, M.P., Price, S.D., Kraemer, K.E., Mizuno, D.R., Carey, S.J., Wright, C.O., Engelke, C.W., Cohen, M., \& Gugliotti, M.G., 
	2003, VizieR On-line Data Catalog, originally published in: Air Force Research Lab. Technical Rep. AFRL-VS-TR-2003-1589 
\bibitem[Fa{\'u}ndez et al., 2004]{faundez_etal04}
	Fa{\'u}ndez, S., Bronfman, L., Garay, G., Chini, R., \& Nyman, L.-\AA., May, J., 2004, A\&A, 426, 97
\bibitem[Fey et al., 1995]{fey_etal95}
	Fey, A.L., Gaume, R.A., Claussen, M.J., \& Vrba, F.J., 1995 ApJ, 453, 308
\bibitem[Finkbeiner et al., 1999]{finkbeiner_etal99}
	Finkbeiner, D.P., Davis, M., \&  Schlegel, D.J., 1999, ApJ, 524, 867
\bibitem[Gehrz et al., 1995]{gehrz_etal95}
	Gehrz, R.D., Hayward, T.L., Houck, J.R., Miles, J.W., Hjellming, R.M., Jones, T.J.,
	Woodward, C.E., Prentice, R., Forrest, W.J., Libonate, S., \& Solomon, S., 1995, ApJ, 439, 417
\bibitem[Giveon et al., 2005a]{giveon_etal05}
	Giveon, U., Becker, R.H., Helfand, D.J., \& White, R.L., 2005a, AJ, 129, 348
\bibitem[Giveon et al., 2005b]{giveon_etal05b}
	Giveon, U., Becker, R.H., Helfand, D.J., \& White, R.L., 2005b, AJ, 130, 156
\bibitem[Haslam et al., 1982]{haslam_etal82}
	Haslam, C.G.T., Salter, C.J., Stoffel, H., \&  Wilson, W.E., 1982, A\&AS, 47, 1
\bibitem[Helfand et al., 2006]{helfand_etal06}
	Helfand, D.J., Becker, R.H., White, R.L., Fallon, A., \& Tuttle, S., 2006, AJ, 131, 2525
\bibitem[Helou \& Walker, 1988]{helou_&_walker88} 
	Helou, G. \& Walker, D.W., 1988, IRAS Catalogs and Atlases Explanatory Supplement, NASA RP-1190, vol 1
\bibitem[Hoare et al., 1991]{hoare_etal91}
	Hoare, M.G., Roche, P.F., Glencross, W.M., 1991, MNRAS, 251, 584
\bibitem[Hunter et al., 2000]{hunter_etal00}
	Hunter, T.R., Churchwell, E., Watson, C., Cox, P., Benford, D.J., \& Roelfsema, P.R., 2000, AJ, 119, 2711
\bibitem[Kurtz et al., 1994]{kurtz_etal94}
	Kurtz, S., Churchwell, E., \& Wood, D.O.S., 1994, ApJS, 91, 659
\bibitem[Launhardt \& Henning, 1997]{launhardt_&_henning97}
	Launhardt, R. \& Henning, T., 1997, A\&A, 326, 329
\bibitem[Leach et al., 2008]{leach_etal08}
	Leach, S.M., Cardoso, J.-F., Baccigalupi, C., Barreiro, R.B., Betoule, M., Bobin, J., 
	Bonaldi, A., Delabrouille, J., de Zotti, G., Dickinson, C., and 20 coauthors 2008, A\&A, 491, 597
\bibitem[Longmore et al., 2007]{longmore_etal07}
	Longmore, S.N., Burton, M.G., Barnes, P.J., Wong, T., Purcell, C.R., \& Ott, J., 2007, MNRAS, 379, 535
\bibitem[Martin-Hernandez et al., 2003]{martin_etal03}
	Martin-Hernandez, N.L., Van Der Hulst, J.M., \& Tielens, A.G.G.M., 2003, A\&A, 407, 957
\bibitem[Motte et al., 2003]{motte_etal03}
	Motte, F., Schilke, P., \& Lis, D.C., 2003, ApJ, 582, 277
\bibitem[Orfei et al., 2002]{orfei_etal02}
	Orfei, A., Morsiani, M., Zacchiroli, G., Maccaferri, G., Roda, J., Fiocchi, F.,
	2002, Proceedings of the 6th European VLBI Network Symposium, 13
\bibitem[Orfei et al., 2004]{orfei_etal04}
	Orfei, A., Morsiani, M., Zacchiroli, G., Maccaferri, G., Roda, J., Fiocchi, F.,
	2004, SPIE, 5495, 116
\bibitem[Ott et al., 1994]{ott_etal94}  
        Ott, M., Witzel, A., Quirrenbach, A., Krichbaum, T.P., Standke, K.J., Schalinski, C.J., \& Hummel, C.A., 1994, A\&A, 284, 3310
\bibitem[Schlegel et al., 1998]{schlegel_etal98}
	Schlegel, D.J., Finkbeiner, D.P., Davis, M., 1998, ApJ, 500, 525
\bibitem[Terzian \& Dickey, 1973]{terzian_&_dickey73}
	Terzian, Y., Dickey, J., 1973, AJ, 78, 875
\bibitem[The Planck Collaboration, 2005]{Planck_col05}
	The Planck Collaboration 2005, ESA-SCI (2005), arXiv:astro-ph/0604069
\bibitem[Thompson et al., 2006]{thompson_etal06}
	Thompson, M.A., Hatchell, J., Walsh, A.J., MacDonald, G.H., \& Millar, T.J., 2006, A\&A, 453, 1003	
\bibitem[Toffolatti et al., 1998]{toffolatti_etal98}
        Toffolatti, L., Argueso Gomez, F., de Zotti, G., Mazzei, P., Franceschini, A., Danese, L., \& Burigana, C., 1998, MNRAS, 297, 117
\bibitem[Umana et al., 2006]{umana_etal06}
        Umana, G., Burigana, C., \& Trigilio, C., 2006, MSAIS, 9, 279
\bibitem[Umana et al., 2008a]{umana_etal08} 
	Umana, G., Leto, P., Trigilio, C., Buemi, C.S., Manzitto, P., Toscano, S., Dolei, S., \& Cerrigone, L.,
	2008a, A\&A, 482, 2, 529
\bibitem[Umana et al., 2008b]{umana_etal08b} 
	Umana, G., Trigilio, C., Cerrigone, L., Buemi, C.S., \& Leto, P.
	2008b, MNRAS, 386, 1404
\bibitem[Walsh et al., 1998]{walsh_etal98}
	Walsh, A.J., Burton, M.G., Hyland, A.R., \& Robinson, G., 1998, MNRAS, 301, 640
\bibitem[White et al., 2005]{white_etal05}
	White, R.L., Becker, R.H., \& Helfand, D.J., 2005, AJ, 130, 586
\bibitem[Williams et al., 2004]{williams_etal04}
	Williams, S.J., Fuller, G.A., \& Sridharan, T.K., 2004, A\&A, 417, 115
\bibitem[Wink et al., 1982]{wink_etal82}
	Wink, J.E., Altenhoff, W.J., \& Mezger, P.G., 1982, A\&A, 108, 227
\bibitem[Wood et al., 1988a]{wood_etal88a}
	Wood, D.O.S., Churchwell, E., \& Salter, C.J., 1988a, ApJ, 325, 694
\bibitem[Wood et al., 1988b]{wood_etal88b} 
	Wood, D.O.S., Handa, T., Fukui, Y., Churchwell, E., Sofue, Y., \& Iwata, T., 1988b, ApJ, 326, 884
\bibitem[Wood \& Churchwell, 1989a]{wood_&_churchwell89a}
	Wood, D.O.S. \& Churchwell, E., 1989a, ApJS, 69, 831
\bibitem[Wood \& Churchwell, 1989b]{wood_&_churchwell89b}
	Wood, D.O.S., Churchwell, E., 1989b, ApJ, 340, 265
\bibitem[Wu et al., 2005]{wu_etal05}
	Wu, Y., Zhu, M., Wei, Y., Xu, D., Zhang, Q., \& Fiege, J.D., 2005, ApJ, 628, L57
\bibitem[Zijlstra et al., 2008]{zijlstra_etal08}	
	Zijlstra, A.A., van Hoof, P.A.M., \& Pearly, R.A., 2008, ApJ, 681, 1296
	
	
%________________________________________________________________
\end{thebibliography}
{}

\end{document}